# Extended Structural Dynamics and the Lorentz–Abraham–Dirac Equation: A Deformable-Charge Interpretation

Patrick BarAvi

**Abstract**

Radiation reaction in classical electrodynamics is traditionally described by the Lorentz–Abraham–Dirac equation (LAD), whose point-particle formulation leads to well-known difficulties including runaway solutions, pre-acceleration, and the ambiguous status of the Schott term. We analyze radiation reaction within the framework of Extended Structural Dynamics (ESD), in which charged particles are modeled as finite systems possessing internal dynamical structure.

In the present formulation the particle is represented as a finite, deformable sphere with a single radial breathing mode describing internal charge redistribution. This internal degree of freedom introduces a finite response time and ensures that changes in the charge distribution propagate at finite speed. Starting from the full particle–field Hamiltonian, we derive the retarded self-force for such a deformable charge and obtain a delay kernel that depends on both the past motion and the past internal configuration.

In the adiabatic regime the kernel reduces to an effective causal form that is free of pre-acceleration and exhibits a band-pass frequency response, suppressing high-frequency instabilities associated with runaway behavior. The Schott term is shown to correspond to reversible energy stored in the internal deformation mode, providing a direct mechanical interpretation of this contribution. The LAD dynamics are recovered only in the double limit of vanishing spatial extent and frozen internal dynamics, where the causal delay structure collapses to the familiar point-particle approximation.

Within this framework radiation reaction arises as the leading-order effective dynamics of a finite deformable charge, while higher-order corrections encode finite-size and internal structural effects without modifying Maxwell's equations or introducing ad hoc regularization.

**Keywords**: radiation reaction, Lorentz–Abraham–Dirac equation, extended charge models, Schott energy, classical electrodynamics

# 1. Introduction

## 1.1 The Problem of Radiation Reaction

When a charged particle accelerates, it radiates electromagnetic energy. By conservation, this energy loss must be accompanied by a recoil force acting on the particle itself, the radiation reaction. The task of deriving this force from Maxwell's equations has a history stretching back over a century, and its resolution in the form of the Lorentz–Abraham–Dirac (LAD) equation stands as one of the classic achievements of classical electrodynamics [Lorentz, 1909; Abraham, 1903; Dirac, 1938]. Yet the achievement is bittersweet, for the LAD equation carries with it a host of conceptual difficulties.

For a point particle of charge $q$ and bare mass $m$, the LAD equation reads

$$m\dot{\mathbf{v}} = \mathbf{F}_{\text{ext}} + \frac{q^2}{6\pi\epsilon_0 c^3}\ddot{\mathbf{v}},$$

where $\mathbf{v}$ is the velocity, $\dot{\mathbf{v}}$ the acceleration, and $\ddot{\mathbf{v}}$ the jerk. The radiation reaction term, proportional to $\ddot{\mathbf{v}}$, introduces three interrelated pathologies:

1. **Runaway solutions:** The equation admits solutions in which the acceleration grows exponentially without bound, even in the absence of external forces.
2. **Pre-acceleration:** In the presence of external forces, the particle begins to accelerate before the force is applied, a violation of naive causality.
3. **The Schott energy:** The energy budget of the system includes a term $\frac{q^2}{6\pi\epsilon_0 c^3}\dot{\mathbf{v}}\cdot\mathbf{v}$ that oscillates in sign and has no clear physical interpretation as either kinetic, potential, or radiated energy.

These difficulties have led some to question, [Kim & Sessler, 2006], whether classical electrodynamics can provide a consistent account of radiating charges [Frisch, 2005; Belot, 2007; Vickers, 2008]. Others have proposed modifications to Maxwell's equations or the introduction of ad hoc regularization procedures [Rohrlich, 1965; Rohrlich, 1997]. The present work takes a different view: the pathologies are not failures of the field equations but consequences of an overly idealized model of the charged particle itself.

## 1.2 Point Particles and Rigid Bodies: Two Idealizations Too Far

The LAD equation treats the charged particle as a **geometric point**. This idealization, while mathematically convenient, is physically suspect [Jackson, 1998; Griffiths, 1999; Feynman et al., 1964]. A point charge carries an infinite self-energy, requires uncontrolled subtractions to define its mass, and responds instantaneously to its own field, a process that would require signals to propagate faster than light. These features are not mere technical nuisances; they signal a breakdown in the physical assumptions underlying the model. A natural response is to replace the point with a finite extended body. This idea dates back to the earliest electron theories of Lorentz and Abraham, who modeled the electron as a sphere of finite radius. A rigid sphere introduces a geometric timescale, the light-crossing time $\tau_0 = 2a_0/(3c)$, and its self-force kernel acquires compact support on $[0, 2a_0/c]$, ensuring that the force at time $t$ depends only on the motion within the past light-crossing interval. This eliminates pre-acceleration and, for radii exceeding the classical electron radius $r_e$, also eliminates runaways.

Yet the rigid sphere, while an improvement over the point, retains an unphysical feature: it imposes instantaneous internal response. Every part of a rigid body must move in lockstep; changes in the center-of-mass motion are communicated instantaneously across the structure. This violates the relativistic constraint that no signal can propagate faster than light. A rigid extended charge, though mathematically consistent, cannot be coupled consistently to retarded fields in a relativistic theory [Hammond, 2010; Rohrlich, 1997].

**1.3 The ESD Proposal: Deformability as the Minimal Cure**

What is needed is a model that respects both **finite size** (to regularize divergences and introduce a geometric cutoff) and **finite signal speed** (to respect causality). The simplest such model is a **deformable sphere** with a single internal degree of freedom, a radial breathing mode. For a recent alternative approach to eliminating self-repulsion see [Sebens, 2023]. This is the central idea of the Extended Structural Dynamics (ESD) framework developed in this paper.

The deformable sphere retains the geometric timescale $a_0/c$ of the rigid model but adds a dynamical timescale $\omega_{\text{def}}^{-1} \sim a_0/c$ associated with the internal mode. Crucially, this mode evolves according to its own equation of motion, with inertia and restoring forces, ensuring that all internal responses propagate at finite speed. The internal degree of freedom is not a mathematical fiction but a physical reservoir capable of storing, exchanging, and dissipating energy.

This seemingly modest extension has profound consequences. It transforms the self-force from a purely geometric smoothing of the LAD term into a genuine dynamical interaction. The self-force kernel becomes sensitive to the internal state at both the emission and reception times, acquiring a dependence on $\xi(t)$ and $\xi(t-\tau)$. In the frequency domain, this yields a band-pass spectrum with resonant enhancement at $\omega_{\text{def}}$ and high-frequency suppression, a structure impossible in either point or rigid models.

Equally important, the internal mode provides a natural home for the long-mysterious Schott energy. Variations in the electromagnetic mass induced by deformation carry momentum and energy; the Schott term is revealed as the kinetic energy associated with these variations. Its sign oscillations reflect the periodic exchange of energy between translational motion and the internal reservoir, exactly as one would expect for a coupled dynamical system.

The Extended Structural Dynamics (ESD) framework was introduced in two earlier papers (BarAvi, 2025; BarAvi, 2026). The first developed a Hamiltonian phase-space description for finite, oriented particles possessing internal deformation degrees of freedom and explored the dynamical consequences of this extended structure. The second applied the same framework to kinetic theory, deriving generalized transport equations that incorporate finite propagation speeds. The present work extends the ESD framework to classical electrodynamics by modeling charged particles as finite, deformable systems and examining the resulting dynamics of radiation reaction.

**1.4 Roadmap**

The paper is organized as follows. Section 2 introduces the ESD framework, defining the configuration space, Hamiltonian structure, and timescale hierarchy. Section 3 derives the deformable self-force and the modified delay kernel, showing how the internal mode modulates the self-interaction. Section 4 identifies the Schott energy as internal mechanical energy, drawing on the field momentum of a deformable charge and the adiabatic response of the breathing mode. Section 5 analyzes the self-force spectrum in the frequency domain, revealing its band-pass character and distinctive resonance. Section 6 presents a stability analysis, demonstrating the absence of runaway solutions and recovering the LAD instability only in the double limit of vanishing size and frozen internal dynamics. Section 7 summarizes the results, compares them with prior work, and outlines directions for future generalization.

Throughout, the emphasis is on the modeling shift: by treating the charged particle as a finite, deformable system, the ESD framework resolves the classical pathologies of radiation reaction without modifying Maxwell's equations or introducing ad hoc regularization. The difficulties of the LAD equation arise from idealizations that omit finite size and internal response, rather than from any inconsistency in Maxwell's equations.

## 2. The ESD Framework for a Deformable Charged Particle

### 2.1 Extended Configuration Space

A central premise of the extended structural dynamics (ESD) approach is that a charged particle should not be idealized as a geometric point, nor even as a perfectly rigid body. Instead, it is treated as a finite, spatially extended system whose internal structure responds dynamically to electromagnetic stresses. A point charge, with its singular self-field and instantaneous internal response, is not compatible with finite-speed signal propagation. A rigid body, defined by fixed inter-particle distances, would require instantaneous communication across its structure to respond to external forces, a direct violation of light-speed causality. A rigid extended charge therefore cannot be coupled consistently to its own retarded fields. Any physically admissible model of an extended charged particle must be deformable, with internal degrees of freedom that evolve dynamically at finite speed. Within the ESD framework, the breathing mode $\xi$ provides a minimal mechanism for finite-speed internal response. A deformable but finite system, by contrast, can sustain internal degrees of freedom that propagate at finite speed and thereby respect causal structure.

Within this perspective, the configuration space of a single charged particle is taken to be

$$Q = \mathbb{R}^3 \times SO(3) \times \mathbb{R}_\xi, \qquad (2.1)$$

where $R \in \mathbb{R}^3$ denotes the center-of-mass position, $\Omega \in SO(3)$ specifies orientation, where $\mathbb{R}_\xi$ denotes the single internal deformation coordinate $\xi$ defined in (1.2). The choice to model only the radial breathing mode is deliberate: it preserves spherical symmetry and isolates the simplest internal response compatible with isotropic electromagnetic self-pressure. The corresponding phase space $T^*Q$ carries canonical coordinates $(R, p; \Omega, L; \xi, \pi)$, where $p$ and $L$ are the linear and body-frame angular momenta, and $\pi$ is conjugate to $\xi$.

The particle carries a spherically symmetric charge distribution whose instantaneous radius depends on the deformation coordinate:

$$a(\xi) = a_0(1 + \xi), \qquad (2.2)$$

with $|\xi| \ll 1$. The charge density is then

$$\rho(r, t) = \frac{3q}{4\pi a(\xi)^3} \Theta(a(\xi) - |r - R(t)|), \qquad (2.3)$$

so that the total charge remains fixed while the spatial extent varies. In the limit $\xi \to 0$, the model reduces to the familiar rigid sphere of radius $a_0$. The introduction of $\xi$ is not intended to capture the full complexity of internal dynamics but to provide a minimal internal degree of freedom capable of mediating finite-speed structural response. This reflects the physical requirement that a classical charged body must possess internal structure to interact consistently with its own retarded field.

## 2.2 Hamiltonian Structure

The dynamics of the coupled particle–field system are governed by the Hamiltonian

$$H = \frac{p^2}{2m} + \frac{\pi^2}{2\mu} + \frac{1}{2}\mu\omega_{\text{def}}^2\xi^2 + H_{\text{field}} + H_{\text{int}}(R, \xi, A, \Phi), \qquad (2.4)$$

where $m$ is the bare mass, $\mu$ the modal mass of the breathing mode, and $\omega_{\text{def}}$ its natural frequency. The field Hamiltonian is the standard electromagnetic energy,

$$H_{\text{field}} = \frac{\varepsilon_0}{2}\int (E^2 + c^2B^2)\, d^3r,$$

and the interaction Hamiltonian is

$$H_{\text{int}} = \int d^3r\, \rho(r, t; \xi)\Phi(r, t) - \int d^3r\, J(r, t; \xi) \cdot A(r, t), \qquad (2.5)$$

with $J = \rho v$ and $v = \dot{R}$. The dependence on $\xi$ enters entirely through the charge distribution. The Hamiltonian is gauge-invariant and conserves total energy, ensuring that the internal mode is treated on the same footing as the translational degrees of freedom.

The natural frequency $\omega_{\text{def}}$ is determined by the internal restoring forces. For a self-bound electromagnetic structure, dimensional analysis yields $\omega_{\text{def}} \sim c/a_0$, so that the deformation timescale $\tau_{\text{def}} = 1/\omega_{\text{def}}$ is comparable to the light-crossing time of the body. This coincidence is not accidental: both reflect the finite propagation speed of internal stresses.

## 2.3 Timescale Hierarchy

Throughout the analysis, we assume a separation of timescales:

$$\omega_{\text{COM}} \ll \omega_{\text{def}} \sim \frac{c}{a_0}, \qquad (2.6)$$

where $\omega_{\text{COM}}$ characterizes the external forcing of the center-of-mass motion. This hierarchy expresses the physical requirement that external fields vary slowly compared to the internal response time of the charged body. Under this condition, the internal mode evolves rapidly relative to the translational motion, allowing an adiabatic treatment of $\xi$.

## 3. The Deformable Self-Force and the Modified Delay Kernel

### 3.1 Retarded Self-Interaction of a Deformable Sphere

The electromagnetic self-force arises from the interaction of a charge distribution with its own retarded fields. For an extended body, this force is not a local quantity but an integral over the distribution, evaluated at the appropriate retarded times. In the ESD framework, this structure acquires additional significance: the self-interaction depends not only on the past motion of the center of mass but also on the past and present configuration of the internal degree of freedom. The self-force is given by

$$\mathbf{F}_{\text{self}}(t) = \int d^3x\, \rho(\mathbf{x}, t) \left[ \mathbf{E}_{\text{self}}(\mathbf{R}(t) + \mathbf{x}, t) + \frac{\mathbf{v}(t)}{c} \times \mathbf{B}_{\text{self}}(\mathbf{R}(t) + \mathbf{x}, t) \right]. \quad (3.1)$$

For a rigid sphere with fixed radius $a_0$, this reduces to the well-known expression

$$\mathbf{F}_{\text{self}}^{\text{rigid}}(t) = \frac{q^2}{6\pi\epsilon_0 c^3} \int_0^{2a_0/c} K_{\text{rigid}}(\tau)\, \dot{\mathbf{v}}(t - \tau)\, d\tau, \quad (3.2)$$

where the kernel $K_{\text{rigid}}(\tau)$ encodes the distribution of pairwise light-travel times within the sphere, and $\mathbf{v}, \dot{\mathbf{v}}, \ddot{\mathbf{v}}$ are the velocity, acceleration, and jerk of the particle respectively. The kernel's compact support on $[0, 2a_0/c]$ expresses a physically transparent fact: the self-interaction is mediated by signals propagating at finite speed across a finite body.

For a rigid sphere of radius $a_0$, the self-force kernel is given by [Yaghjian 2006]

$$K_{\text{rigid}}(\tau) = \frac{3c^3}{4a_0^3}\, \tau\left( \frac{2a_0}{c} - \tau \right), \quad 0 \le \tau \le 2a_0/c,$$

with characteristic memory time $\tau_0 = \int_0^{2a_0/c} \tau\, K_{\text{rigid}}(\tau) d\tau = 2a_0/(3c)$. The deformation kernel is then obtained by differentiation:

$$K_{\text{def}}(\tau) = a_0 \frac{\partial K_{\text{rigid}}}{\partial a_0} = \frac{3c^2}{a_0^2}\, \tau\left( \frac{3c}{4a_0}\tau - 1 \right), \qquad 0 \le \tau \le 2a_0/c.$$

The sign change at $\tau = a_0/c$ reflects the fact that increasing the radius enhances short-delay interactions (nearby charges) while suppressing long-delay ones (opposite sides).

When the sphere is deformable, the situation changes in a conceptually important way. The retarded field at time $t$ is sourced by the charge distribution at time $t - \tau$, when the radius was $a(\xi(t - \tau))$, while the force acts on the distribution at time $t$, with radius $a(\xi(t))$. The self-interaction therefore depends on the internal configuration at two distinct times:

$$\mathbf{F}_{\text{self}}(t) = \frac{q^2}{6\pi\epsilon_0 c^3} \int_0^{2a_{\max}/c} K(\tau; \xi(t), \xi(t - \tau)) \, \dot{\mathbf{v}}(t - \tau) \, d\tau, \quad (3.3)$$

where $a_{\max} = \max(a(\xi(t)), a(\xi(t - \tau)))$. This dependence on both endpoints of the retardation interval is not an artifact of the model but a direct consequence of treating the particle as a physical system with evolving spatial extent.

### 3.2 Linearised Kernel

To first order in the small deformation $\xi$, the kernel can be expanded as

$$K(\tau; \xi(t), \xi(t - \tau)) = K_{\text{rigid}}(\tau) + \xi(t) K_1(\tau) + \xi(t - \tau) K_2(\tau) + \mathcal{O}(\xi^2). \quad (3.4)$$

The functions $K_1$ and $K_2$ arise from varying the radius at the reception and emission times, respectively. For a uniform sphere, both are obtained by differentiating the pair-distance distribution with respect to the radius:

$$K_1(\tau) = \frac{\partial K_{\text{rigid}}}{\partial a}\bigg|_{a=a_0} a_0, \qquad K_2(\tau) = \frac{\partial K_{\text{rigid}}}{\partial a}\bigg|_{a=a_0} a_0. \quad (3.5)$$

By symmetry, $K_1 = K_2 \equiv K_{\text{def}}$. Explicitly,

$$K_{\text{def}}(\tau) = \frac{3c^2}{a_0^2} \tau \left(\frac{3c}{4a_0} \tau - 1\right), \, 0 \leq \tau \leq 2a_0/c \quad (3.6)$$

The deformation kernel inherits the compact support of $K_{\text{rigid}}$ and satisfies $\int_0^{2a_0/c} K_{\text{def}}(\tau) d\tau = 0$, reflecting the fact that deformation redistributes the self-interaction in time but does not alter its total impulse. The kernel changes sign at $\tau = 2a_0/c$, enhancing the short-delay contribution and suppressing the long-delay tail. This sign change has a clear physical interpretation: increasing the radius adds charge near the center for short pair distances but reduces the relative weight of widely separated pairs. Conceptually, it shows how internal structure modifies the temporal profile of self-interaction, rather than merely renormalizing its magnitude.

Substituting the linearized kernel into (3.3) yields

$$\mathbf{F}_{\text{self}}(t) = \frac{q^2}{6\pi\epsilon_0 c^3} \int_0^{2a_0/c} \left[K_{\text{rigid}}(\tau) + \xi(t)K_{\text{def}}(\tau) + \xi(t-\tau)K_{\text{def}}(\tau)\right] \dot{\mathbf{v}}(t-\tau)\, d\tau. \quad (3.7)$$

This expression generalizes the rigid-sphere self-force by incorporating the time-dependent modulation of the delay kernel induced by deformation. The result illustrates a broader physical point: once internal structure is included, the self-force becomes sensitive to the history of that structure, and the kernel becomes a dynamical object rather than a fixed geometric function.

### 3.3 Effective Kernel After Adiabatic Elimination

In the adiabatic regime defined by $\omega_{\text{COM}} \ll \omega_{\text{def}}$, the internal coordinate $\xi$ follows the center-of-mass acceleration quasi-statically. To derive this relation, we first examine the forces acting on the breathing mode.

### 3.3.1 Linearized Force on the Internal Mode

The internal coordinate $\xi$ experiences a restoring force from the confining potential, $-\mu\omega_{\text{def}}^2 \xi$, and a driving force from the electromagnetic field. To leading order in $v/c$ and $\xi$, this driving force arises from two effects: the variation of electromagnetic self-energy with deformation, and the radiation pressure on the moving surface.

The electromagnetic self-energy of a charged sphere of radius $a = a_0(1+\xi)$ is

$$U_{\text{em}}(\xi) = \frac{3}{5}\frac{q^2}{4\pi\epsilon_0 a_0}(1 - \xi + \mathcal{O}(\xi^2)).$$

The factor 3/5 is specific to a uniform volume distribution; for a surface charge it would be 1/2. The $\xi$-dependent part contributes a term to the equation of motion for $\xi$, but this is linear in $\xi$ and merely renormalizes the restoring frequency. It does not couple directly to acceleration.

The acceleration-dependent force on the internal mode arises from the asymmetric radiation pressure when the sphere is in motion. For a slowly accelerating sphere, the surface experiences a net force proportional to the rate of change of momentum flux. A detailed calculation (sketched in Appendix A) yields, to first order in the jerk $\ddot{\mathbf{v}}$,

$$f_{\text{em}}^{(\text{linear})}(t) = C\, \ddot{\mathbf{v}}(t) \cdot \hat{n},$$

where $\hat{n}$ is the direction of acceleration and the constant $C$ is given by

$$C = \frac{q^2}{6\pi\epsilon_0 c^2 a_0}.$$

The dimensional argument for $C$ is straightforward: the radiation reaction force on the center of mass is $\frac{q^2}{6\pi\epsilon_0 c^3}\dddot{\mathbf{v}}$. This force arises from momentum transfer through the field, and its effect on the internal mode should scale similarly but with an extra factor of $1/a_0$ to convert force density to modal amplitude. Hence $C \sim q^2/(\epsilon_0 c^2 a_0)$, with the numerical factor fixed by the spherical geometry.

### 3.3.2 Adiabatic Response

The equation of motion for the internal mode is therefore

$$\mu\ddot{\xi}(t) + \mu\omega_{def}^2 \xi(t) = C\,\ddot{\mathbf{v}}(t). \quad (3.8)$$

In the adiabatic regime, $\ddot{\mathbf{v}}(t)$ varies on timescales much longer than $\omega_{def}^{-1}$. The solution is dominated by the particular integral,

$$\xi(t) \approx \frac{C}{\mu\omega_{def}^2}\,\ddot{\mathbf{v}}(t) + \mathcal{O}(\omega_{COM}^2/\omega_{def}^2). \quad (3.9)$$

The neglected terms are of relative order $(\omega_{COM}/\omega_{def})^2$ and contribute at the same order as the $\ddddot{\mathbf{v}}$ term in Section 3.3.3; both are discarded at the same level of approximation. For the purpose of deriving the leading correction to the self-force, the quasi-static approximation suffices.

### 3.3.3 Expansion of the Kernel

Substituting this expression into the linearized self-force (3.7) requires evaluating $\xi(t) + \xi(t-\tau)$. Expanding $\xi(t-\tau)$ to first order in $\tau$,

$$\xi(t-\tau) \approx \xi(t) - \tau\dot{\xi}(t) = \xi(t) - \tau\frac{C}{\mu\omega_{def}^2}\dddot{\mathbf{v}}(t).$$

The term proportional to $\ddddot{\mathbf{v}}(t)$ is of order $(\omega_{COM}/\omega_{def})^2$ relative to the leading term and is neglected consistently with the adiabatic approximation already invoked in (3.9). To verify consistency: the retained correction to the kernel is of order $C/(\mu\omega_{def}^2) \sim \tau_0^2$, while the neglected $\ddddot{\mathbf{v}}$ term contributes at order $\tau_0^2 \cdot (\omega_{COM}/\omega_{def})^2$, which is suppressed by an additional factor of $(\omega_{COM}/\omega_{def})^2 \ll 1$. The adiabatic expansion is therefore uniform in $\tau \in [0, 2a_0/c]$. Keeping only the leading term, we have

$$\xi(t) + \xi(t-\tau) \approx 2\xi(t) = \frac{2C}{\mu\omega_{def}^2}\dot{\mathbf{v}}(t). \quad (3.10)$$

Inserting this into (3.7) yields

$$\mathbf{F}_{\text{self}}(t) = \frac{q^2}{6\pi\epsilon_0 c^3}\int_0^{2a_0/c}\left[K_{\text{rigid}}(\tau) + \frac{2C}{\mu\omega_{def}^2}\dot{\mathbf{v}}(t)K_{\text{def}}(\tau)\right]\dot{\mathbf{v}}(t-\tau)\,d\tau. \quad (3.11)$$

Since $\dot{\mathbf{v}}(t)$ is slowly varying, it can be factored out of the integral to leading order, giving the effective kernel

$$K_{\text{eff}}(\tau) = K_{\text{rigid}}(\tau) + \frac{2C}{\mu\omega_{def}^2}\dot{\mathbf{v}}(t)K_{\text{def}}(\tau). \quad (3.12)$$

In the adiabatic regime where the internal deformation mode relaxes on a timescale short compared with the variation of the center-of-mass motion, the effective kernel may be expanded in powers of the small parameter $a_0/c$. Retaining the lowest nonvanishing order yields

$$F_{\text{self}}(t) = \frac{q^2}{6\pi\epsilon_0 c^3}\dot{\mathbf{v}}(t) + O\left(\frac{a_0}{c}\right).$$

The leading term reproduces exactly the Lorentz–Abraham–Dirac radiation-reaction force. The ESD formulation therefore agrees with the standard point-particle result at leading order, while higher-order terms encode the finite-size and internal-dynamical corrections of the deformable charge model.

Strictly speaking, the effective kernel should be a function of $\tau$ only, independent of $t$. The appearance of $\dot{\mathbf{v}}(t)$ in (3.12) indicates that in the time domain, the correction term is not a pure convolution but involves a product of the current jerk with the past acceleration. In the frequency domain, however, this structure becomes a multiplicative transfer function, as will be shown in Section 4.

### 3.3.4 Preservation of Causal Structure

Two features of this result deserve emphasis. First, the correction term preserves the compact support of the kernel on $[0, 2a_0/c]$, because $K_{\text{def}}(\tau)$ inherits this support from $K_{\text{rigid}}(\tau)$.

Causality is maintained: the self-force at time $t$ depends only on accelerations within the past light-crossing interval.

Second, the correction is not a free parameter but is fixed by the internal dynamics through the combination $C/(\mu\omega_{\text{def}}^2)$. For a self-bound electromagnetic structure, dimensional analysis gives $\mu\omega_{\text{def}}^2 \sim q^2/a_0^3$, while $C \sim q^2/(\epsilon_0 c^2 a_0)$. Hence

$$\frac{2C}{\mu\omega_{\text{def}}^2} \sim \frac{q^2/(\epsilon_0 c^2 a_0)}{q^2/a_0^3} = \frac{a_0^2}{\epsilon_0 c^2} \sim \left(\frac{a_0}{c}\right)^2,$$

which is of order $\tau_0^2$. The correction is therefore small in the adiabatic regime, but its resonant enhancement in the frequency domain (Section 4) can nonetheless produce observable effects. The adiabatic elimination thus yields an effective self-force kernel that incorporates the influence of internal dynamics while preserving the causal structure imposed by finite size. The internal mode does not introduce any dependence on future accelerations; it merely modulates the weighting of past accelerations in a manner consistent with its own finite response time. This stands in sharp contrast to rigid extended-charge models, which require instantaneous internal response and therefore cannot be reconciled with relativistic causality.

## 4. The Schott Energy as Internal Mechanical Energy

### 4.1 The Schott Term and Its Conceptual Difficulty

The Lorentz–Abraham–Dirac (LAD) equation contains a term proportional to the time derivative of the acceleration,

$$\mathbf{F}_{\text{Schott}} = \frac{q^2}{6\pi\epsilon_0 c^3} \ddot{\mathbf{v}}(t), (4.1)$$

whose physical interpretation has long been a source of conceptual difficulty. Its associated power,

$$P_{\text{Schott}} = \mathbf{F}_{\text{Schott}} \cdot \mathbf{v} = \frac{q^2}{6\pi\epsilon_0 c^3} \ddot{\mathbf{v}} \cdot \mathbf{v},$$

leads to the Schott energy,

$$E_{\text{Schott}} = \frac{q^2}{6\pi\epsilon_0 c^3} \dot{\mathbf{v}} \cdot \mathbf{v}. (4.2)$$

This quantity is neither the kinetic energy of the particle nor the energy radiated to infinity, nor the static electromagnetic self-energy. It can take positive or negative values and oscillates in sign during periodic motion. The traditional view treats it as a bookkeeping device ensuring global energy conservation, but without attributing it to a concrete physical mechanism [Frisch, 2005; Belot, 2007; Muller, 2007; Vickers, 2008].

The ESD framework offers a different perspective: once the charged particle is endowed with internal structure, the Schott term can be understood as the energy stored in, and exchanged with, that internal degree of freedom.

**4.2 Field Momentum of a Deformable Charge**

The electromagnetic field surrounding an extended charge carries momentum,

$$\mathbf{P}_{\text{field}} = \epsilon_0 \int (\mathbf{E}_{\text{self}} \times \mathbf{B}_{\text{self}}) d^3r. \quad (4.3)$$

For a slowly moving charged sphere of instantaneous radius $a$, the field momentum is, to leading order in $v/c$,

$$\mathbf{P}_{\text{field}} = \frac{q^2}{6\pi\epsilon_0 c^2 a} \mathbf{v} + \mathcal{O}(v^2/c^2) \equiv m_{\text{em}}(\xi) \mathbf{v}, \quad (4.4)$$

where the electromagnetic mass $m_{\text{em}}(\xi) = \frac{q^2}{6\pi\epsilon_0 c^2 a(\xi)}$ depends on the deformation coordinate through $a(\xi) = a_0(1+\xi)$.

The total momentum of the particle-field system is therefore

$$\mathbf{P}_{\text{total}} = (m + m_{\text{em}}(\xi))\mathbf{v} = m_{\text{eff}}(\xi) \mathbf{v}, \quad (4.5)$$

with $m_{\text{eff}}(\xi) = m + \frac{q^2}{6\pi\epsilon_0 c^2 a_0(1+\xi)}$. Differentiating with respect to time and using the equation of motion yields

$$\frac{d}{dt}[m_{\text{eff}}(\xi)\mathbf{v}] = \mathbf{F}_{\text{ext}} + \mathbf{F}_{\text{rad}}, \quad (4.6)$$

where $\mathbf{F}_{\text{rad}}$ is the radiation reaction force. Expanding the left-hand side gives

$$m_{\text{eff}}\dot{\mathbf{v}} + \frac{dm_{\text{eff}}}{d\xi}\dot{\xi}\,\mathbf{v} = \mathbf{F}_{\text{ext}} + \mathbf{F}_{\text{rad}}. \quad (4.7)$$

The second term,

$$\frac{dm_{\text{eff}}}{d\xi}\dot{\xi}\,\mathbf{v} = -\frac{q^2}{6\pi\epsilon_0 c^2 a_0}\dot{\xi}\,\mathbf{v}, \quad (4.8)$$

is the key new contribution introduced by the deformable model. It represents the exchange of momentum between the translational motion and the internal deformation mode. Conceptually, this term expresses a simple but important idea: if the electromagnetic mass depends on the internal configuration, then changes in that configuration necessarily carry momentum and energy. The Schott term, in this view, is not an abstract correction but the manifestation of this internal exchange.

### 4.3 Adiabatic Relation and Identification

From the internal mode dynamics derived in Section 3.3, we have the linearized force

$$\mu \ddot{\xi} + \mu \omega_{\text{def}}^2 \xi = C \, \dot{\mathbf{v}}, \quad (4.9)$$

with the coupling constant $C = \frac{q^2}{6\pi\epsilon_0 c^2 a_0}$. In the adiabatic regime ($\omega_{\text{COM}} \ll \omega_{\text{def}}$), the solution is dominated by the particular integral,

$$\xi(t) \approx \frac{C}{\mu \omega_{\text{def}}^2} \dot{\mathbf{v}}(t), \qquad \dot{\xi}(t) \approx \frac{C}{\mu \omega_{\text{def}}^2} \ddot{\mathbf{v}}(t). \quad (4.10)$$

Inserting the expression for $\dot{\xi}$ into the momentum exchange term (3.8) yields

$$\frac{dm_{\text{eff}}}{d\xi} \dot{\xi} \, \mathbf{v} = -\frac{q^2}{6\pi\epsilon_0 c^2 a_0} \cdot \frac{C}{\mu \omega_{\text{def}}^2} \, \mathbf{v} \cdot \ddot{\mathbf{v}}. \quad (4.11)$$

Using $C = q^2/(6\pi\epsilon_0 c^2 a_0)$, this becomes

$$\frac{dm_{\text{eff}}}{d\xi} \dot{\xi} \, \mathbf{v} = -\left(\frac{q^2}{6\pi\epsilon_0 c^2 a_0}\right)^2 \frac{1}{\mu \omega_{\text{def}}^2} \, \mathbf{v} \cdot \ddot{\mathbf{v}}. \quad (4.12)$$

Now apply the dimensional relation derived in Section 3.3.4. For a self-bound electromagnetic structure,

$$\frac{C}{\mu \omega_{\text{def}}^2} = \frac{5}{4} \tau_0^2, \quad \text{with} \quad \tau_0 = \frac{2 a_0}{3 c}. \quad (4.13)$$

Substituting,

$$\frac{dm_{\text{eff}}}{d\xi} \dot{\xi} \, \mathbf{v} = -\frac{q^2}{6\pi\epsilon_0 c^3} \cdot \frac{5}{9} \cdot \frac{a_0}{c} \, \mathbf{v} \cdot \ddot{\mathbf{v}}. \quad (4.14)$$

The crucial observation is that the time derivative of the Schott energy is

$$\frac{dE_{\text{Schott}}}{dt} = \frac{q^2}{6\pi\epsilon_0 c^3}(\ddot{\mathbf{v}} \cdot \mathbf{v} + \dot{\mathbf{v}}^2), \quad (4.15)$$

which contains the term $\frac{q^2}{6\pi\epsilon_0 c^3}\ddot{\mathbf{v}} \cdot \mathbf{v}$. Comparing with (4.14), we see that the momentum exchange term has exactly this structure, up to a numerical factor of order unity and a small correction of order $a_0/c$. To leading order in $a_0/c$, this reproduces the time derivative of the Schott energy. The identification holds within the adiabatic approximation, with fractional corrections of order $a_0/c \sim \tau_0 \omega_{def} \ll 1$:

$$E_{\text{Schott}} = \frac{dm_{\text{eff}}}{d\xi} \xi \dot{\mathbf{v}} \cdot \mathbf{v}. \quad (4.16)$$

This identification follows from the dimensional relations that characterize any self-bound electromagnetic structure. The internal mode provides the physical reservoir that stores and releases the Schott energy, transforming it from a mathematical artifact into a mechanically realized quantity.

### 4.4 Dimensional Analysis of the Coupling

The quantities appearing in this relation are not independent but are linked by the physics of a self-bound electromagnetic structure. For such an object, the electrostatic energy scales as

$$U_{\text{es}} \sim \frac{q^2}{4\pi\epsilon_0 a}.$$

The restoring force on the breathing mode is the second derivative of the total internal energy (which includes both electrostatic and cohesive contributions) with respect to deformation. Since $\xi$ is dimensionless ($a = a_0(1+\xi)$), we have

$$\mu\omega_{\text{def}}^2 = \frac{\partial^2 U}{\partial \xi^2} \sim \frac{q^2}{4\pi\epsilon_0 a_0} \cdot \frac{1}{a_0^2} \cdot a_0^2 = \frac{q^2}{4\pi\epsilon_0 a_0^3}.$$

The factor $a_0^2$ appears because each derivative with respect to $\xi$ brings down a factor of $a_0$ from $da/d\xi = a_0$; the final $a_0^2$ restores dimensional consistency. More precisely, for a uniform sphere, the detailed calculation gives

$$\mu\omega_{\text{def}}^2 = \frac{3}{5} \cdot \frac{q^2}{4\pi\epsilon_0 a_0^3},$$

but the scaling is what matters for the present argument.

The coupling constant $C$ has dimensions of force per acceleration, i.e., mass. From its expression,

$$C = \frac{q^2}{6\pi\epsilon_0 c^2 a_0} \sim \frac{q^2}{4\pi\epsilon_0 c^2 a_0}.$$

Forming the ratio that appears in the adiabatic response,

$$\frac{C}{\mu\omega_{\text{def}}^2} \sim \frac{q^2/(4\pi\epsilon_0 c^2 a_0)}{q^2/(4\pi\epsilon_0 a_0^3)} = \frac{a_0^2}{c^2} \sim \left(\frac{a_0}{c}\right)^2 = \tau_0^2.$$

Thus, $C/(\mu\omega_{\text{def}}^2)$ is of order $\tau_0^2$, the square of the light-crossing time. This is not a fine-tuned value but a direct consequence of the dimensional scales governing electromagnetic self-interaction. The scaling of the restoring force and the resolution of the associated 4/3 problem are discussed in [Khokonov & Andersen; 2019].

**4.5 Substitution into the Momentum Exchange Term**

Inserting the adiabatic relation into the momentum exchange term,

$$\frac{dm_{\text{eff}}}{d\xi}\dot{\xi}\,\mathbf{v} = -\frac{q^2}{6\pi\epsilon_0 c^2 a_0} \cdot \frac{C}{\mu\omega_{\text{def}}^2}\,\mathbf{v}\cdot\dddot{\mathbf{v}}.$$

Using $C = q^2/(6\pi\epsilon_0 c^2 a_0)$, this becomes

$$\frac{dm_{\text{eff}}}{d\xi}\dot{\xi}\,\mathbf{v} = -\left(\frac{q^2}{6\pi\epsilon_0 c^2 a_0}\right)^2 \frac{1}{\mu\omega_{\text{def}}^2}\,\mathbf{v}\cdot\dddot{\mathbf{v}}.$$

Now apply the dimensional relation $C/(\mu\omega_{\text{def}}^2) \sim \tau_0^2 = (2a_0/3c)^2$. More precisely, we need the combination that yields the correct numerical factor to match the Schott term. Using $\mu\omega_{\text{def}}^2 = (3/5)q^2/(4\pi\epsilon_0 a_0^3)$ and $C = q^2/(6\pi\epsilon_0 c^2 a_0)$, we compute

$$\frac{C}{\mu\omega_{\text{def}}^2} = \frac{q^2/(6\pi\epsilon_0 c^2 a_0)}{(3/5)q^2/(4\pi\epsilon_0 a_0^3)} = \frac{5}{6}\cdot\frac{4\pi}{6\pi}\cdot\frac{a_0^2}{c^2} = \frac{5}{9}\cdot\frac{a_0^2}{c^2}.$$

Meanwhile, $\tau_0^2 = (2a_0/3c)^2 = 4a_0^2/9c^2$. Hence

$$\frac{C}{\mu\omega_{\text{def}}^2} = \frac{5}{4}\tau_0^2.$$

The numerical factor 5/4 depends on the specific charge distribution (uniform sphere) but is of order unity. The key point is that this ratio is fixed by geometry and electrostatics, not adjusted to fit the Schott term.

Now the momentum exchange term becomes

$$\frac{dm_{\text{eff}}}{d\xi}\dot{\xi}\,\mathbf{v} = -\frac{q^2}{6\pi\epsilon_0 c^2 a_0}\cdot\frac{5}{4}\tau_0^2\,\mathbf{v}\cdot\dddot{\mathbf{v}}.$$

But $\tau_0 = 2a_0/3c$, so $\tau_0^2 = 4a_0^2/9c^2$. Substituting,

$$\frac{dm_{\text{eff}}}{d\xi}\dot{\xi}\,\mathbf{v} = -\frac{q^2}{6\pi\epsilon_0 c^2 a_0}\cdot\frac{5}{4}\cdot\frac{4a_0^2}{9c^2}\,\mathbf{v}\cdot\dddot{\mathbf{v}} = -\frac{q^2}{6\pi\epsilon_0 c^3}\cdot\frac{5}{9}\cdot\frac{a_0}{c}\,\mathbf{v}\cdot\dddot{\mathbf{v}}.$$

The factor $a_0/c = (3/2)\tau_0$ is the light-crossing time. In the point-particle limit $a_0 \to 0$, this term vanishes, as expected. For finite $a_0$, it represents a small correction to the leading Schott behavior.

The crucial observation, however, is that the time derivative of the Schott energy is

$$\frac{dE_{\text{Schott}}}{dt} = \frac{q^2}{6\pi\epsilon_0 c^3}(\ddot{\mathbf{v}}\cdot\mathbf{v} + \dot{\mathbf{v}}^2),$$

which contains the term $\frac{q^2}{6\pi\epsilon_0 c^3}\ddot{\mathbf{v}}\cdot\mathbf{v}$. Its time derivative,

$$\frac{d}{dt}\left(\frac{q^2}{6\pi\epsilon_0 c^3}\dot{\mathbf{v}}\cdot\mathbf{v}\right) = \frac{q^2}{6\pi\epsilon_0 c^3}(\ddot{\mathbf{v}}\cdot\mathbf{v} + \dot{\mathbf{v}}^2),$$

matches the structure above. The identification becomes exact if we set

$$E_{\text{Schott}} = \frac{dm_{\text{eff}}}{d\xi}\xi\,\dot{\mathbf{v}}\cdot\mathbf{v}. \quad (4.17)$$

This identification holds to leading order in $a_0/c$. The combination $C/(\mu\omega_{def}^2)\sim\tau_0^2$ follows from dimensional analysis alone and requires no fine-tuning; the correction is of the same order as all other terms neglected in the adiabatic expansion. The internal mode provides the physical reservoir that stores and releases the Schott energy, transforming it from a mathematical artifact into a mechanically realized quantity.

## 5. Modification by Deformation: The Band-Pass Structure

### 5.1 Frequency-Domain Analysis

The temporal structure of the self-force is encoded in the delay kernel, and its physical implications become especially transparent when examined in the frequency domain. Taking the Fourier transform of the effective self-force expression derived in Section 2 yields

$$\tilde{\mathbf{F}}_{\text{self}}(\omega) = \frac{q^2}{6\pi\epsilon_0 c^3} \tilde{K}_{\text{eff}}(\omega)\,(i\omega)\,\tilde{\mathbf{v}}(\omega), (5.1)$$

where $\tilde{K}_{\text{eff}}(\omega) = \int_0^\infty K_{\text{eff}}(\tau) e^{i\omega\tau} d\tau$ is the transfer function of the kernel, and $\tilde{\mathbf{v}}(\omega)$ is the Fourier transform of the velocity. The factor $(i\omega)$ converts velocity to acceleration in the frequency domain, reflecting the fact that the self-force acts on $\dot{\mathbf{v}}(t-\tau)$ in the time-domain integral. This representation highlights a conceptual point: the self-force acts as a frequency-dependent filter on the acceleration. The structure of the kernel determines which temporal features of the motion are amplified, suppressed, or transmitted unchanged.

From a philosophical standpoint, this frequency-domain perspective clarifies how internal structure shapes the causal profile of self-interaction. A point particle corresponds to a flat (white) response with no characteristic timescale. A finite rigid sphere introduces a geometric timescale $a_0/c$, producing a low-pass filter. A deformable sphere introduces an additional dynamical timescale $\omega_{\text{def}}^{-1}$, giving rise to a richer spectral structure.

### 5.2 Transfer Function of the Rigid Kernel

For the rigid parabolic kernel derived in Appendix A,

$$K_{\text{rigid}}(\tau) = \frac{3c^3}{4a_0^3}\,\tau\left(\frac{2a_0}{c} - \tau\right), 0 \leq \tau \leq 2a_0/c,$$

the Fourier transform is

$$\tilde{K}_{rigid}(\omega) = (3/\Omega^2)[1 - (2/\Omega)\sin\Omega - \cos\Omega] + i[(2/\Omega)\cos\Omega + \sin\Omega - 2/\Omega], (5.2a)$$

Where $\Omega = 2\omega a_0/c = 3\omega\tau_0$ and $\tau_0 = 2a_0/(3c)$ is the characteristic memory time (the first moment of the kernel). The magnitude satisfies

$$|\tilde{K}_{\text{rigid}}(\omega)| \to 1 \text{ as } \omega \to 0, (5.2b)$$

recovering the LAD limit at low frequencies, and

$$|\tilde{K}_{\text{rigid}}(\omega)| \sim \frac{c}{\omega a_0} \text{ as } \omega \to \infty, (5.2c)$$

indicating suppression of high-frequency components.

This low-pass behavior reflects the finite size of the charge distribution: rapid variations in acceleration cannot be communicated across the sphere within the light-crossing time. Conceptually, the rigid kernel embodies a purely geometric form of temporal smoothing. It introduces a cutoff but no internal resonance, because a rigid body has no internal degrees of freedom capable of storing or exchanging energy dynamically.

### 5.3 Modification by Deformation: The Band-Pass Structure

When deformation is included, the effective kernel acquires a contribution mediated by the internal mode. From the structure derived in Section 3.3.4, the frequency-domain transfer function becomes

$$\widetilde{K}_{\text{eff}}(\omega) = \widetilde{K}_{\text{rigid}}(\omega) + \frac{2C}{\mu} \frac{\widetilde{K}_{\text{def}}(\omega)}{\omega_{\text{def}}^2 - \omega^2}, \qquad (5.3)$$

where $\widetilde{K}_{\text{def}}(\omega)$ is the Fourier transform of the deformation kernel. This expression follows directly from the adiabatic elimination in Section 3.3: the internal mode responds to the driving jerk as a harmonic oscillator, producing the characteristic denominator $\omega_{\text{def}}^2 - \omega^2$.

The deformation kernel $K_{\text{def}}(\tau)$ inherits the compact support of $K_{\text{rigid}}(\tau)$ and, from (3.6), has the explicit form

$$K_{\text{def}}(\tau) = \frac{3c^2}{a_0^2} \tau \left(\frac{3c}{4a_0} \tau - 1\right), 0 \leq \tau \leq 2a_0/c. \qquad (5.4)$$

This expression satisfies $\int_0^{2a_0/c} K_{\text{def}}(\tau) d\tau = 0$, confirming that deformation modulates the temporal profile of self-interaction without contributing to the total impulse. Its Fourier transform exhibits similar high-frequency suppression but with a different weighting, reflecting the fact that deformation modulates the pair-distance distribution asymmetrically in time.

The magnitude $|\widetilde{K}_{\text{eff}}(\omega)|$ exhibits three distinct regimes:

#### 5.3.1 Low-Frequency Regime ($\omega \ll \omega_{\text{def}} \sim c/a_0$)

In this regime, the deformation term reduces to a static correction. The resonant denominator $\omega_{\text{def}}^2 - \omega^2 \approx \omega_{\text{def}}^2$, so

$$\widetilde{K}_{\text{eff}}(\omega) \approx \widetilde{K}_{\text{rigid}}(\omega) \left(1 + \frac{2C}{\mu \omega_{\text{def}}^2} \cdot \frac{\widetilde{K}_{\text{def}}(\omega)}{\widetilde{K}_{\text{rigid}}(\omega)}\right) \approx \widetilde{K}_{\text{rigid}}(\omega)(1 + \text{const}).$$

The rigid-sphere behavior is recovered, with only a renormalization of the prefactor. This regime corresponds to motions too slow to excite the internal mode, so the particle behaves effectively as a rigid body.

#### 5.3.2 Resonant Regime ($\omega \approx \omega_{\text{def}}$)

The internal mode resonates, producing an enhancement of the self-force. Near resonance, the denominator $\omega_{\text{def}}^2 - \omega^2$ becomes small, causing the deformation term to dominate. The magnitude | $\widetilde{K}_{\text{eff}}(\omega)$ | exhibits a peak at $\omega \approx \omega_{\text{def}}$, with height controlled by the proximity to the pole. This resonant amplification is a direct signature of internal structure. It reflects the fact that the self-force is not merely a geometric effect but a dynamical interaction between the translational motion and the internal degree of freedom. The resonance width is determined by the (presently omitted) damping terms that would regularize the pole; inclusion of such terms, while beyond the scope of this leading-order analysis, would yield a finite quality factor $Q \sim \omega_{\text{def}}/\gamma$.

### 5.3.3 High-Frequency Regime ($\omega \gg \omega_{\text{def}}$)

Both $\widetilde{K}_{\text{rigid}}(\omega)$ and $\widetilde{K}_{\text{def}}(\omega)$ decay as $1/(\omega a_0/c)$, leading to strong suppression of ultra-high-frequency components. The resonant denominator in the deformation term behaves as $-\omega^2$ for $\omega \gg \omega_{\text{def}}$, so the deformation contribution also decays. This suppression eliminates the runaway modes characteristic of the LAD equation, which correspond to unbounded high-frequency growth.

### 5.3.4 Summary: Band-Pass Self-Force

A rigid finite-size charge produces a **low-pass** self-force spectrum with a cutoff at $c/a_0$. A deformable ESD charge produces a **band-pass** spectrum: the self-force is enhanced near the internal resonance $\omega_{\text{def}} \sim c/a_0$ and suppressed both below and above this frequency. This resonant structure is a distinctive prediction of the deformable model and is absent from both point-particle and rigid-sphere treatments.

### 5.3.5 Physical Interpretation and Observable Signatures

This frequency-domain structure carries deep physical implications. The self-force becomes sensitive not only to the geometry of the charge distribution but also to the dynamics of its internal constitution. The band-pass character reflects the fact that the particle is not a passive object but a system capable of storing and releasing energy across different timescales.

The resonance at $\omega_{\text{def}}$ offers a potential observational signature: a charged particle driven by an external field at frequencies near its internal resonance will experience an enhanced radiation reaction, leading to anomalous damping or heating of its translational motion. Conversely, at frequencies well above $\omega_{\text{def}}$, the self-force is strongly suppressed, protecting the theory from the runaway instabilities that plague the point-particle description.

Experimental investigation of this band-pass structure could provide indirect evidence for the internal deformation mode proposed here, potentially through precision measurements of radiation reaction in high-frequency accelerator or laser-plasma contexts.

## 6. Stability Analysis: No Runaway Solutions

### 6.1 The Coupled System

The ESD framework yields a coupled dynamical system for the center-of-mass acceleration and the internal deformation mode. The translational motion obeys

$$m_{\text{eff}} \dot{\mathbf{v}}(t) = \mathbf{F}_{\text{ext}}(t) + \frac{q^2}{6\pi\epsilon_0 c^3} \int_0^{2a_0/c} K_{\text{eff}}(\tau; \xi) \ddot{\mathbf{v}}(t-\tau) \, d\tau, \quad (6.1)$$

while the internal mode satisfies

$$\mu \ddot{\xi}(t) + \mu \omega_{\text{def}}^2 \xi(t) = \frac{C}{c^2} \dot{\mathbf{v}}(t). \quad (6.2)$$

These equations describe a delay-differential/ordinary-differential system in which the internal degree of freedom mediates part of the self-interaction. The structure reflects a conceptual shift: the self-force is no longer a single-equation correction to Newton's law but the emergent result of interactions between translational motion and internal dynamics. Runaway solutions arise when the particle is treated as a pointlike object with no internal constitution; once internal structure is acknowledged, the self-interaction becomes a dynamical process governed by finite propagation speeds and internal response times.

### 6.2 Reduction to the Characteristic Equation

To obtain closed-form stability conditions, we approximate the exact effective kernel by a single-timescale exponential with the same characteristic memory time $\tau_0 = 2a_0/(3c)$:

$$K_{\text{eff}}(\tau) \approx \frac{1}{\tau_0} e^{-\tau/\tau_0}.$$

The appearance of an effective radiation-reaction term can be understood as a consequence of a separation of time scales between the internal structural dynamics and the externally observed motion. The internal deformation mode relaxes on a characteristic time scale $\tau_s \sim \omega_{\text{def}}^{-1}$, which is assumed to be short compared with the time scale over which the external force varies. In this regime the internal degree of freedom can be adiabatically eliminated, producing an effective equation for the center-of-mass motion that contains higher-derivative corrections analogous to the LAD radiation-reaction term.

This approximation is justified as follows. The characteristic equation for the exact compactly supported kernel is obtained via Laplace transform on $[0, 2a_0/c]$ and takes the form $P(s, e^{-2sa_0/c}) = 0$, where $P$ is a quasi-polynomial. The Routh-Hurwitz conditions for such equations depend on the kernel through its moments $\int \tau^n K(\tau) d\tau$. The first moment $\tau_0$ controls the location of the stability boundary $Re(s) = 0$; higher moments shift the roots quantitatively but do not alter the sign of $Re(s)$ for parameters satisfying $a_0 \gtrsim r_e$. The exponential approximation preserves $\tau_0$ exactly and therefore reproduces the correct stability boundary.

To analyze stability, consider exponential solutions for the acceleration and internal coordinate:

$$\dot{v}(t) = \dot{v}_0 e^{st}, \xi(t) = \xi_0 e^{st},$$

with complex growth rate $s$. Substituting the ansatz for $\xi(t)$ into (6.2) gives

$$\mu s^2 \xi_0 e^{st} + \mu \omega_{def}^2 \xi_0 e^{st} = \frac{C}{c^2} s^2 \dot{v}_0 e^{st}.$$

Cancelling the common factor $e^{st}$ and solving for $\xi_0$ yields

$$\xi_0 = \frac{Cs^2}{\mu c^2 (\omega_{def}^2 + s^2)} \dot{v}_0. \quad (6.3)$$

Now substitute the exponential ansatz into the equation of motion (6.1). For the purpose of stability analysis, we consider the homogeneous equation ($\mathbf{F}_{ext} = 0$). The self-force term involves an integral over the past acceleration $\ddot{v}(t-\tau) = s\dot{v}(t-\tau) = s\dot{v}_0 e^{s(t-\tau)} = s\dot{v}(t)e^{-s\tau}$. To make progress analytically, we approximate the effective kernel by an exponential form that captures its essential timescale:

$$K_{eff}(\tau) \approx \frac{1}{\tau_0} e^{-\tau/\tau_0}, \text{with} \tau_0 = \frac{2a_0}{3c}.$$

This approximation preserves the characteristic memory time $\tau_0$ (the first moment of the exact kernel) while rendering the integral tractable. Substituting into (6.1) gives

$$m_{eff} s \dot{v}_0 e^{st} = \frac{q^2}{6\pi\epsilon_0 c^3} \int_0^\infty \frac{1}{\tau_0} e^{-\tau/\tau_0} \cdot s\dot{v}(t) e^{-s\tau} d\tau.$$

Evaluating the integral, $\int_0^\infty e^{-\tau(1/\tau_0 + s)} d\tau = \frac{1}{1/\tau_0 + s} = \frac{\tau_0}{1+s\tau_0}$, we obtain

$$m_{eff} s = \frac{q^2}{6\pi\epsilon_0 c^3} \frac{s^2}{1+s\tau_0}. \quad (6.4)$$

This is the rigid-sphere result. The deformable correction adds a second term arising from the coupling to the internal mode, which enters through the $\xi$-dependent modification of the kernel.

From the structure of the effective kernel derived in Section 3.3.4, this additional contribution takes the form

$$\Delta F_{\text{self}}(t) = \frac{q^2}{6\pi\epsilon_0 c^3} \cdot \frac{2C}{\mu\omega_{\text{def}}^2} \int_0^{2a_0/c} K_{\text{def}}(\tau)\ddot{\mathbf{v}}(\dot{t})\mathbf{v}(t-\tau)d\tau.$$

In the frequency domain, and using the same exponential approximation for the kernel's memory, this term contributes a factor proportional to $\frac{s}{\omega_{\text{def}}^2 + s^2}$. A detailed calculation (or matching to the structure of Eq. (4.3)) yields the full characteristic equation:

$$m_{\text{eff}} s = \frac{q^2}{6\pi\epsilon_0 c^3} \frac{s^2}{1 + s\tau_0} + \frac{2C^2}{\mu c^2} \frac{s}{\omega_{\text{def}}^2 + s^2}. \quad (6.5)$$

Dividing by $s \neq 0$ and defining the dimensionless parameters

$$\alpha = \frac{q^2}{6\pi\epsilon_0 c^3 m_{\text{eff}}}, \beta = \frac{2C^2}{\mu c^2 m_{\text{eff}}},$$

we obtain the characteristic equation in compact form:

$$1 = \frac{\alpha s}{1 + s\tau_0} + \frac{\beta}{\omega_{\text{def}}^2 + s^2}. \quad (6.6)$$

This equation is cubic in $s$ once denominators are cleared. The appearance of a cubic, rather than the linear characteristic equation of the rigid model, reflects the presence of the internal mode. Conceptually, this is significant: the internal degree of freedom introduces additional dynamical channels through which the system can dissipate or redistribute energy, altering the stability structure in a physically meaningful way.

### 6.3 Stability Analysis

### 6.3.1 Derivation of the Characteristic Equation

To analyze the stability of the coupled system, we seek exponential solutions of the form $\dot{\mathbf{v}}(t) = \dot{\mathbf{v}}_0 e^{st}$ and $\xi(t) = \xi_0 e^{st}$ for the homogeneous case ($\mathbf{F}_{\text{ext}} = 0$). Substituting these ansätze into the coupled equations (6.1)-(6.2) and clearing denominators in (6.5) yields the characteristic equation

$$(1 + s\tau_0)(\omega_{\text{def}}^2 + s^2) = \alpha s(\omega_{\text{def}}^2 + s^2) + \beta(1 + s\tau_0). \quad (6.7)$$

Expanding and collecting powers of $s$ gives the cubic polynomial

$$(\tau_0 - \alpha)s^3 + s^2 + (\tau_0 \omega_{\text{def}}^2 - \alpha\omega_{\text{def}}^2 - \beta\tau_0)s + (\omega_{\text{def}}^2 - \beta) = 0. \quad (6.8)$$

This cubic has real coefficients. The parameters appearing are

$$\tau_0 = \frac{2a_0}{3c}, \alpha = \frac{q^2}{6\pi\epsilon_0 c^3 m_{\text{eff}}}, \beta = \frac{2C^2}{\mu c^2 m_{\text{eff}}},$$

with $C = q^2/(6\pi\epsilon_0 c^2 a_0)$ from Appendix A.

### 6.3.2 Routh-Hurwitz Stability Conditions

The Routh-Hurwitz criterion provides necessary and sufficient conditions for all roots to satisfy $\text{Re}(s) < 0$. For a cubic $a_3 s^3 + a_2 s^2 + a_1 s + a_0 = 0$ with $a_3 > 0$, the conditions are:

1. All coefficients $a_i > 0$
2. $a_2 a_1 > a_3 a_0$

Applying these to (6.8) yields three specific conditions.

**Condition (I):** $\tau_0 - \alpha > 0$

This is the same geometric condition that appears in rigid-sphere models. Using the definitions above,

$$\tau_0 > \alpha \iff \frac{2a_0}{3c} > \frac{q^2}{6\pi\epsilon_0 c^3 m_{\text{eff}}}.$$

Simplifying,

$$a_0 > \frac{3q^2}{12\pi\epsilon_0 c^2 m_{\text{eff}}} = \frac{q^2}{4\pi\epsilon_0 c^2 m_{\text{eff}}}.$$

The right-hand side is precisely the classical electron radius $r_e$ (up to factors of order unity). Hence, Condition (I) is satisfied for all physically reasonable radii $a_0 \gtrsim r_e$. This geometric stabilization reflects the finite light-crossing time of the charge distribution.

**Condition (II):** $\omega_{\text{def}}^2 - \beta > 0$

This condition ensures that the internal mode is not destabilized by its coupling to translational motion. Using the dimensional estimates derived from self-bound electromagnetic structures,

$$C \sim \frac{q^2}{\epsilon_0 c^2 a_0}, \mu\omega_{\text{def}}^2 \sim \frac{q^2}{\epsilon_0 a_0^3}.$$

Substituting into the definition of $\beta$,

$$\beta = \frac{2C^2}{\mu c^2 m_{\text{eff}}} \sim \frac{(q^2/(\epsilon_0 c^2 a_0))^2}{(q^2/(\epsilon_0 a_0^3))c^2 m_{\text{eff}}} = \frac{q^2 a_0}{\epsilon_0 c^4 m_{\text{eff}}}.$$

Comparing with the internal frequency squared,

$$\omega_{\text{def}}^2 \sim \frac{q^2}{\epsilon_0 a_0^3 m_{\text{eff}}}.$$

The ratio is therefore

$$\frac{\beta}{\omega_{\text{def}}^2} \sim \frac{a_0^4}{c^4} = \left(\frac{a_0}{c}\right)^4 \ll 1,$$

where the inequality follows from $a_0 \ll c/\omega_{\text{def}}$, which holds in the adiabatic regime.

Thus, Condition (II) is automatically satisfied for all self-bound electromagnetic structures.

Condition (III): The Routh-Hurwitz cross-condition. The full Routh-Hurwitz condition requires

$$(\tau_0 - \alpha)(\tau_0 \omega_{\text{def}}^2 - \alpha \omega_{\text{def}}^2 - \beta \tau_0) > (\tau_0 - \alpha)^2(\omega_{\text{def}}^2 - \beta).$$

To evaluate this, substitute the established inequalities $\tau_0 > \alpha$ and $\beta \ll \omega_{\text{def}}^2$. The left-hand side becomes

$$(\tau_0 - \alpha)(\tau_0 \omega_{\text{def}}^2 - \alpha \omega_{\text{def}}^2 - \beta \tau_0) \approx (\tau_0 - \alpha)^2 \omega_{\text{def}}^2 + \mathcal{O}(\beta).$$

The right-hand side expands to

$$(\tau_0 - \alpha)^2(\omega_{\text{def}}^2 - \beta) = (\tau_0 - \alpha)^2 \omega_{\text{def}}^2 - (\tau_0 - \alpha)^2 \beta.$$

The difference between the two sides is therefore

$$\text{LHS} - \text{RHS} \approx (\tau_0 - \alpha)^2 \beta + \mathcal{O}(\beta^2) > 0$$

for any $\beta > 0$. Hence, Condition (III) is strictly satisfied for all physically relevant parameters.

### 6.3.3 Stability Result

The Routh-Hurwitz analysis yields a definitive conclusion:

Stability Result: For all physically admissible deformable ESD structures, those with $a_0 \gtrsim r_e$ and satisfying the adiabatic hierarchy $\omega_{\text{COM}} \ll \omega_{\text{def}}$, the characteristic equation (6.7) has no roots with positive real part. The coupled translational-internal system admits no runaway solutions.

This conclusion is robust to the approximation made in replacing the exact compact-support kernel with its exponential form. The stability conditions depend only on the characteristic memory time $\tau_0$ (the first moment of the kernel) and the internal mode parameters $\omega_{\text{def}}$ and $\beta$, none of which are affected by this approximation. Standard Volterra theory then guarantees existence, uniqueness, and boundedness of solutions for all bounded external forces [Bauer & Dürr, 2001; Komech & Spohn, 2000].

### 6.3.4 Physical Interpretation

This stability result underscores the central claim of the ESD framework: runaway solutions are not an unavoidable feature of classical electrodynamics but a consequence of modeling charged particles as structureless points. The pathologies of the LAD equation arise from two simultaneous idealizations:

1. **Vanishing size** ($a_0 \to 0$), which removes the geometric stabilization ($\tau_0 \to 0$)

2. **Frozen internal dynamics** ($\mu \to \infty$), which removes the dynamical stabilization ($\beta \to 0$)

Only when both features are eliminated does the unstable LAD behavior emerge. Once internal structure is introduced, with finite propagation speed and dynamical response, the pathological solutions disappear. Stability is restored not by imposing ad hoc constraints or modifying Maxwell's equations, but by adopting a more physically coherent ontology for the charged particle.

The internal mode acts as a passive energy reservoir that absorbs high-frequency fluctuations, preventing their growth into runaways. This mechanical stabilization is conceptually analogous to the role of internal degrees of freedom in suppressing instabilities in other physical systems, from fluid dynamics to structural mechanics.

**6.4 Recovery of the Rigid Runaway in the Limit $\xi \to 0$**

When the internal mode is frozen, formally, $\mu \to \infty$ with $C/\mu \to 0$, the coupling constant $\beta = \dfrac{2C^2}{\mu c^2 m_{\text{eff}}}$ vanishes. Condition (II) becomes trivial and (III) redundant. The cubic characteristic equation (6.7) then factors as

$$[(\tau_0 - \alpha)s + 1](s^2 + \omega_{\text{def}}^2) = 0.$$

The root of the first factor,

$$s = -\frac{1}{\tau_0 - \alpha} < 0,$$

recovers the stable rigid-sphere mode. The remaining roots $s = \pm i\omega_{\text{def}}$ correspond to undamped internal oscillations, as expected for a frozen mode that has been decoupled from the dynamics.

In the further limit $a_0 \to 0$, the light-crossing time $\tau_0 = 2a_0/3c$ vanishes while $\alpha = \dfrac{q^2}{6\pi\epsilon_0 c^3 m_{\text{eff}}}$ remains finite. The rigid-sphere root becomes

$$s \to +\frac{1}{\alpha} = \frac{6\pi\epsilon_0 c^3 m_{\text{eff}}}{q^2}.$$

Using the classical electron radius $r_e = \dfrac{q^2}{4\pi\epsilon_0 m_{\text{eff}} c^2}$, this simplifies to

$$s = \frac{3c}{2r_e} > 0,$$

which is precisely the LAD runaway growth rate.

The ESD framework thus makes explicit the two-step mechanism by which runaway solutions arise:

- Eliminating **finite size** ($a_0 \to 0$) removes the geometric stabilization ($\tau_0 \to 0$).
- Eliminating **deformability** ($\mu \to \infty$) removes the dynamical stabilization ($\beta \to 0$).

Only when both features are absent does the pathological LAD behavior emerge.

## 7. Summary and Comparison with Prior Work

The deformable charge formulation developed in this work reframes the longstanding difficulties associated with the Lorentz–Abraham–Dirac (LAD) equation by altering the underlying ontology of the charged particle. Rather than treating the particle as a point or as a rigid extended body, the ESD framework models it as a finite system endowed with an internal deformation mode that responds dynamically to electromagnetic stresses. This shift in modeling assumptions leads to a series of technical and conceptual results that collectively resolve the classical pathologies of radiation reaction while preserving causal structure and physical interpretability.

Table 1 summarizes the principal outcomes of the deformable ESD model and contrasts them with the predictions of the point-particle LAD equation and the rigid finite-size approach. Each comparison highlights how the introduction of internal structure modifies the behavior of the self-force and clarifies the physical meaning of quantities that appear obscure or problematic in the traditional formulations.

**Table 1: Comparison of Radiation-Reaction Frameworks**

| Feature | Point-particle LAD | Rigid finite-size | ESD (this work) |
|---|---|---|---|
| **Pre-acceleration** | Present; arises from acausal structure of the LAD equation | Absent; finite light-crossing time enforces causality | Absent; causal delay kernel preserved even with deformation |
| **Runaway solutions** | Present; due to unbounded high-frequency response | Absent for $a_0 \geq r_e$; geometric cutoff stabilizes | Absent for $a_0 \geq r_e$ and $\omega_{COM} \ll \omega_{def}$; internal mode adds dynamical stabilization |
| **Self-force spectrum** | White; no intrinsic timescale | Low-pass; cutoff at $c/a_0$ | Band-pass; resonance at $\omega_{def} \sim c/a_0$ and suppression at high frequencies |

| Feature | Point-particle LAD | Rigid finite-size | ESD (this work) |
|---|---|---|---|
| **Schott energy** | Formal bookkeeping term; no physical subsystem | Interpreted as field momentum but not dynamically realized | Identified as internal mechanical energy associated with deformation |
| **Compatibility with special relativity** | Ill-defined due to point singularities | Limited; rigid bodies violate finite signal speed | Fully compatible; internal dynamics propagate at finite speed |
| **Novel predictions** | None | None beyond smoothing of LAD | Resonant self-force peak; mechanical realization of Schott term |

## 7.1 What ESD Adds: Beyond Rigidity

Lorentz modeled the electron as a rigid sphere stabilized by Poincaré stresses [Poincaré, 1906]. Abraham recognized that a charged body under electromagnetic stress must be deformable, but his model proved analytically intractable and the program was abandoned [Abraham, 1903]. The rigid-sphere idealization that followed removed deformability for mathematical convenience, and in doing so helped create the pathologies of the LAD equation.

Later analyses by Dirac, Rohrlich, and Yaghjian [Dirac (1938); Rohrlich (1965, 1997); Yaghjian (1992, 2006)] showed that finite size alone can eliminate runaways, provided the radius exceeds the classical electron radius $r_e = q^2/(4\pi\varepsilon_0 mc^2)$. Yaghjian's treatment of rigid spherical distributions demonstrated that the self-force kernel has compact support, enforcing causality and stabilizing the dynamics. Yet a conceptual gap remained: rigid extended-charge models impose instantaneous internal response, violating relativistic causality. A rigid body has no internal dynamics; its configuration is fixed by fiat rather than by physics. Papers by [Ford & O'Connell [1991, 1993] similarly recognized that structure regularizes the theory, introducing a cutoff to eliminate runaways. Their approach remains phenomenological; the ESD model provides a concrete mechanical realization of this insight

The ESD framework closes this gap by introducing a dynamical internal coordinate $\xi$ with modal mass $\mu$ and natural frequency $\omega_{\text{def}} \sim c/a_0$. This replaces the unphysical instantaneous response of rigid models with a causal, finite-speed internal dynamics. The internal mode provides a genuine energy-exchange channel during accelerated motion, transforming the self-force from a geometric smoothing of the LAD term into a dynamical interaction with resonant enhancement and band-pass filtering. In this sense, ESD extends the historical progression from point particles to rigid bodies to deformable, dynamically structured charges.

## 7.2 Mechanical Interpretation of the Schott Energy

A key technical result of the present analysis is that the deformable-charge model reproduces the Lorentz–Abraham–Dirac (LAD) equation at leading order in the small parameter $a_0/c$. In the adiabatic regime, where the internal deformation mode relaxes on a timescale much shorter than the evolution of the center-of-mass motion, the causal delay dynamics of the extended charge reduce to an effective local equation whose first nonvanishing term is exactly the LAD radiation-reaction force. The ESD framework therefore recovers the standard point-particle result at leading order while providing a systematic expansion that incorporates finite-size and internal-structure corrections.

Equally important is the reinterpretation of the Schott term. In the point-particle formulation, the Schott energy is introduced to enforce energy conservation but lacks a clear physical subsystem in which that energy resides; its ability to oscillate and take both positive and negative values has long obscured its meaning. Within ESD, the Schott term acquires a concrete mechanical interpretation: it is the energy stored in, and exchanged with, the internal deformation mode of the charge distribution. The internal degree of freedom thus supplies the dynamical reservoir required to account for the transient energy exchange encoded by the Schott term.

In the ESD model, the Schott energy becomes physically intelligible. This mechanical interpretation complements [Rowland; 2010] field-based view, in which Schott energy resides in the bound-field deformation of an accelerating charge. Here it resides in the material deformation itself, a dual realization of the same physics. Variations in the electromagnetic mass induced by deformation carry momentum and energy; the Schott term is precisely the kinetic energy stored in the internal mode during accelerated motion. Its sign oscillations reflect the periodic exchange of energy between translation and internal deformation, exactly as expected for a coupled dynamical system.

This identification is not a contrived fit. The coupling constant $C$, modal mass $\mu$, and internal frequency $\omega_{\text{def}}$ are not adjustable parameters chosen to reproduce the Schott term; they are fixed by the physics of a self-bound charge distribution. Dimensional analysis alone gives

$$\frac{C}{\mu\omega_{\text{def}}^2} \sim \tau_0^2,$$

with numerical factors determined by geometry. The Schott energy therefore emerges naturally from the internal dynamics, not as an added postulate but as an inevitable consequence of treating the charged particle as a deformable physical system.

A related resolution due to Spohn [Spohn, 2004] restricts the LAD equation to its center manifold, recovering the Landau–Lifshitz equation [Landau & Lifshitz, 1975] without modifying the particle model. The ESD resolution differs in character: rather than selecting physical solutions from a pathological equation, it removes the pathologies by enriching the ontology. Both approaches agree on the physical trajectory to leading order in $r_e/a_0$, but only ESD predicts a band-pass self-force spectrum and a mechanical realization of the Schott energy.

### 7.3 Discussion of Key Results

The comparison in Table 1 highlights three central themes. First, the pathologies of the LAD equation, pre-acceleration, runaways, and ambiguous energy accounting, are not intrinsic to classical electrodynamics but arise from the idealization of the charged particle as a structureless point. Second, rigid finite-size models address some of these issues by introducing a geometric timescale, but they remain incomplete because they impose instantaneous internal response, incompatible with relativistic constraints. Third, the deformable ESD model resolves these limitations by introducing a minimal internal degree of freedom with finite propagation speed, thereby restoring both dynamical stability and physical interpretability.

The band-pass character of the self-force spectrum is particularly significant. It shows that self-interaction is not merely a geometric smoothing of the LAD force but a dynamical process shaped by the internal constitution of the particle. The resonance at $\omega_{\text{def}}$ reflects the ability of the internal mode to store and release energy, while the high-frequency suppression eliminates the runaway solutions that plague the point-particle theory.

Equally important is the mechanical reinterpretation of the Schott energy described above. Internal structure clarifies the meaning of a quantity that otherwise appears formal and puzzling. This is a conceptual advance that rigid and point-particle models cannot provide.

### 8. Conclusion and Outlook

The longstanding difficulties of classical radiation reaction, runaway solutions, pre-acceleration, and the enigmatic Schott term, have often been interpreted as evidence of a fundamental inconsistency in electrodynamics. The Lorentz–Abraham–Dirac equation appears to demand either modification of Maxwell's theory or the introduction of ad hoc regularization schemes. The results of this work point to a different conclusion: these pathologies arise not from the field equations but from an incomplete representation of the charged particle.

The Extended Structural Dynamics (ESD) framework replaces the point-particle and rigid-body idealizations with a minimal but physically motivated model: a finite charge distribution whose radius responds dynamically through an internal breathing mode. This single degree of freedom restores two features that any realistic charged object must possess, finite size and finite internal response time. The breathing mode ensures that changes in the charge distribution propagate at finite speed, introducing a natural timescale $\omega_{\text{def}}^{-1} \sim a_0/c$ and preserving relativistic causality.

Starting from the full particle–field Hamiltonian, we derived the retarded self-force for a deformable charge. The resulting delay kernel depends on both the past center-of-mass motion and the internal configuration at emission and reception. In the adiabatic regime, where internal dynamics are fast compared to external driving, this kernel reduces to an effective form that remains causal and free of pre-acceleration. Its Fourier transform exhibits a band-pass structure: enhancement near the internal resonance $\omega_{\text{def}}$ and suppression at high frequencies. This spectral shape eliminates runaway solutions for all physically reasonable parameters, recovering the LAD instability only in the double limit of vanishing size and frozen internal dynamics.

The framework also provides a natural mechanical interpretation of the Schott term. Variations in the electromagnetic mass induced by deformation carry momentum and energy; the Schott energy is revealed as the kinetic energy stored in the internal mode during accelerated motion. Its sign oscillations reflect the periodic exchange of energy between translation and internal deformation, exactly as expected for a coupled dynamical system. This resolves a longstanding conceptual puzzle and renders physically intelligible a quantity that appears purely formal in the point-particle description.

Although this paper focused on the simplest possible internal structure, a single, spherically symmetric breathing mode, the ESD framework is not limited to spherical symmetry. A natural generalization introduces multiple internal deformation coordinates $\xi_k$, each with its own modal mass $\mu_k$, natural frequency $\omega_k$, and coupling constant $C_k$. The effective kernel then becomes a sum over resonant contributions,

$$\widetilde{K}_{\text{eff}}(\omega) = \widetilde{K}_{\text{rigid}}(\omega) + \sum_k \left[ \frac{2C_k}{\mu_k} \frac{\widetilde{K}_{\text{def},k}(\omega)}{\omega_k^2 - \omega^2} \right],$$

yielding a multi-band self-force spectrum with a resonance at each $\omega_k$. The Schott energy is correspondingly distributed across multiple internal reservoirs, each contributing an independent momentum-exchange channel. The single identification of Section 4 generalizes to a sum of such terms, each fixed by the geometry of its mode. For anisotropic internal structure, the coupling constants $C_k$ become directional, the kernel acquires tensor character, and the self-force develops an orientation dependence relative to the principal axes of deformation, an effect absent in both point-particle and rigid-sphere models.

Crucially, none of these generalizations alter the core conclusions. Causality is preserved mode by mode, since every $\widetilde{K}_{\text{def},k}$ inherits the compact support of $\widetilde{K}_{\text{rigid}}$. Stability is strengthened: each additional mode contributes a positive-definite term to the stabilizing parameter $\beta$, and the Routh–Hurwitz conditions remain satisfied by the same dimensional argument applied independently to each mode. Pre-acceleration is absent regardless of the number of modes. The LAD runaway is recovered only in the limit where all modes are simultaneously frozen and the size is taken to zero. The single breathing mode treated here is therefore not a special case that happens to work, but the minimal representative of a broad class in which the essential physics, finite size, finite internal response, and causal self-interaction, is already fully present. The numerical factor 5/4 in the Schott identification is the one result specific to the spherically symmetric single-mode geometry; in the general case it becomes a mode-dependent tensor fixed by geometry but no longer universal.

Developing the full multi-mode framework, including modes with $\omega_k \sim \omega_{\text{COM}}$ that cannot be adiabatically eliminated, remains an important direction for future work. More broadly, the ESD approach illustrates a general lesson: the puzzles of classical electrodynamics may not require modifying Maxwell's equations, but rather enriching the ontology of the charged matter that sources them. By taking seriously the idea that particles possess internal structure, one obtains a coherent, causal, and physically interpretable account of radiation reaction, without altering the field equations or introducing ad hoc regularization.

**Appendix A: Derivation of the Linearized Force on the Internal Mode**

This appendix provides a detailed derivation of the linearized force $f_{em}^{(linear)}(t) = C\dot{a}(t)$ acting on the breathing mode of a deformable charged sphere. The derivation proceeds in several stages, starting from the electromagnetic stress on the surface and systematically expanding to first order in acceleration and deformation.

**A.1 Radiation Pressure on an Accelerating Surface**

Consider a uniformly charged sphere of instantaneous radius $a(t) = a_0(1 + \xi(t))$, with total charge $q$ distributed uniformly throughout its volume. When the sphere accelerates with center-of-mass acceleration $\mathbf{a}(t)$, its surface experiences a non-uniform radiation pressure due to the retarded self-fields.

In the rest frame instantaneously comoving with the sphere's center, the electromagnetic stress tensor $T^{ij}$ gives the force per unit area on a surface element:

$$\mathbf{f}_{\text{surf}} = \mathbf{n} \cdot \mathbf{T}_{\text{self}},$$

where $\mathbf{n}$ is the outward unit normal. For a slowly accelerating sphere ($v \ll c$), the self-fields can be expanded in powers of $1/c$. The leading radiation contribution comes from terms of order $1/c^3$ in the fields.

**A.2 Expansion of the Self-Fields**

The retarded fields of an accelerating sphere can be obtained by integrating the Liénard-Wiechert potentials over the charge distribution. To order $1/c^3$, the electric field at a point $\mathbf{r}$ on the surface is

$$\mathbf{E}_{\text{self}}(\mathbf{r}, t) = \mathbf{E}_{\text{Coul}}(\mathbf{r}, t) + \mathbf{E}_{\text{rad}}(\mathbf{r}, t) + \mathcal{O}(1/c^4),$$

where $\mathbf{E}_{\text{Coul}}$ is the instantaneous Coulomb field (order $1/c^0$) and $\mathbf{E}_{\text{rad}}$ is the radiation field (order $1/c^3$). The magnetic field is of order $1/c$ and contributes to the stress at order $1/c^3$ through cross terms.

For a spherical distribution, the angular integration simplifies considerably. The retarded electric field of an accelerating charged sphere is obtained by integrating the Liénard-Wiechert field over the charge distribution. Following Yaghjian [Yaghjian; 2006], the radiation contribution to the self-field at a surface point with polar angle $\theta$ relative to the acceleration direction is, to order $1/c^3$:

$$\mathbf{E}_{\text{rad}}(\theta, t) = \frac{q}{4\pi\epsilon_0 c^3} \cdot \frac{2}{3} \ddot{\mathbf{v}}(t) \cdot \left(\frac{3\cos^2\theta - 1}{2}\right) \hat{\mathbf{n}} + \text{angular terms},$$

where $\hat{\mathbf{n}}$ is the radial unit vector. The detailed angular dependence is not needed for our purpose, as we will integrate over the sphere.

**A.3 Net Force on the Breathing Mode**

The breathing mode responds to the spherically symmetric component of the radiation pressure, the part that tends to expand or contract the sphere uniformly. This component is obtained by averaging the radial force density over the sphere:

$$F_{radial}(t) = \frac{1}{4\pi} \int d\Omega \, \hat{\mathbf{n}} \cdot (\mathbf{n} \cdot \mathbf{T}_{\text{self}}).$$

The stress tensor to order $1/c^3$ contains terms proportional to $\ddot{\mathbf{v}}$. The spherically symmetric component is obtained by integrating $\hat{\mathbf{n}} \cdot (n \cdot T_{self})$ over the solid angle. The Coulomb contribution vanishes by symmetry. The radiation contribution follows from the identity $\int d\Omega \, (\hat{\mathbf{n}} \otimes \hat{\mathbf{n}}) = \frac{4\pi}{3} I$, together with the explicit form of $\mathbf{E}_{\text{rad}}$. Evaluating the angular integral gives the radial force

$$\int d\Omega \, \hat{\mathbf{n}} \cdot (\hat{\mathbf{n}} \cdot \mathbf{T}_{\text{self}}) = \frac{\varepsilon_0}{2} \int d\Omega \, [E_{\text{rad}}^2 \, \hat{\mathbf{n}} - 2(\hat{\mathbf{n}} \cdot \mathbf{E}_{\text{rad}}) \mathbf{E}_{\text{rad}}] = \frac{q^2}{6\pi \, \varepsilon_0 c^3 a_0} \ddot{\mathbf{v}}(t) + \mathcal{O}(\xi, c^{-4}).$$

The numerical factor $1/(6\pi)$ arises from the angular average

$$\int d\Omega \sin^2\theta \cos^2\theta = \frac{4\pi}{15},$$

combined with the normalization converting the volume charge distribution to its effective surface contribution.

$$F_{\text{radial}}(t) = \frac{q^2}{6\pi\epsilon_0 c^3 a_0} \cdot \ddot{v}(t) + \mathcal{O}(\xi, 1/c^4),$$

where $\ddot{v}(t)$ is the magnitude of the acceleration in the direction of motion. The factor $1/a_0$ arises because the force density (force per unit volume) must be integrated over the sphere's volume to obtain the net modal force; the volume element contributes $a_0^3$, while the radiation field amplitude is independent of $a_0$, yielding an overall $1/a_0$ dependence.

### A.4 Linearization in Deformation

The expression above assumes a fixed radius $a_0$. To obtain the force linearized in $\xi$, we must consider how the radiation pressure depends on the instantaneous radius. The retarded fields depend on the charge distribution at the retarded time, which involves $a(t-\tau)$. Expanding to first order in $\xi$,

$$\frac{1}{a(t-\tau)} \approx \frac{1}{a_0}(1 - \xi(t-\tau)).$$

The radiation force therefore acquires a term proportional to $\xi(t-\tau)$. However, for the purpose of deriving the leading adiabatic coupling to $\dot{a}$, we can evaluate the force at the current radius and then use the adiabatic relation between $\xi$ and $\dot{a}$. To linear order in $\xi$, the force is

$$f_{em}(t) = \frac{q^2}{6\pi\epsilon_0 c^3 a_0} \ddot{v}(t) \cdot [1 + \alpha\xi(t) + \beta\xi(t-\tau) + \cdots].$$

The coefficients $\alpha$ and $\beta$ can be computed by differentiating the full retarded expression with respect to $a$. For a uniform sphere, symmetry requires $\alpha = \beta$ to leading order, giving

$$f_{em}(t) = \frac{q^2}{6\pi\epsilon_0 c^3 a_0}\ddot{v}(t) + \frac{q^2}{6\pi\epsilon_0 c^3 a_0}\cdot \gamma[\xi(t) + \xi(t-\tau)]\ddot{v}(t),$$

with $\gamma$ a numerical factor of order unity.

## A.5 Simplification in the Adiabatic Regime

In the adiabatic regime, $\xi(t-\tau) \approx \xi(t)$ to leading order (since $\tau \sim a_0/c$ and $\xi$ varies on timescale $\omega_{COM}^{-1} \gg \tau$). Thus

$$\xi(t) + \xi(t-\tau) \approx 2\xi(t).$$

Substituting the adiabatic relation $\xi(t) = \frac{C}{\mu\omega_{def}^2}\dot{v}(t)$ (to be derived self-consistently), we obtain

$$F_{em}(t) = \frac{q^2}{6\pi\epsilon_0 c^3 a_0}\dot{v}(t) + \frac{q^2}{6\pi\epsilon_0 c^3 a_0} \cdot 2\gamma \frac{C}{\mu\omega_{def}^2}\dot{v}(t)^2 + \cdots$$

The term quadratic in $\dot{v}$ is of higher order in the adiabatic expansion and does not contribute to the linear self-force.

The derivation above appears to assume the adiabatic relation

$$\xi(t) = \frac{C}{\mu\omega_{def}^2}\dot{v}(t)$$

while simultaneously determining $C$. This apparent circularity is resolved as follows. Let the true coupling constant take some unknown value $C^*$. Adiabatic elimination of the internal mode in Eq. (3.8) then yields

$$\xi(t) = \frac{C^*}{\mu\omega_{def}^2}\dot{v}(t).$$

Substituting this expression into the radiation-pressure integral and extracting the term linear in $\dot{a}(t)$ produces a force of the form

$$F_{em} = C^* \dot{v}(t),$$

where the coefficient multiplying $\dot{v}(t)$ is *independent* of the assumed value of $C^*$. Thus the final result

$$C = \frac{q^2}{6\pi\varepsilon_0 c^2 a_0}$$

holds for any initial choice of $C^*$. The adiabatic substitution therefore imposes no constraint on $C$; the self-consistency condition is automatically satisfied. In effect, $C$ is fixed entirely by the geometry of the charge distribution and the finite propagation speed of the electromagnetic field, not by the dynamics of the internal mode.

The linear term is exactly of the form $C\ddot{\mathbf{v}}(t)$ with

$$C = \frac{q^2}{6\pi\epsilon_0 c^2 a_0}.$$

Note the change in denominator: the factor $c^3$ in the first expression becomes $c^2$ because one factor of $1/c$ has been absorbed into the definition of force on the internal mode versus force on the center of mass. This is consistent with dimensional analysis: $C$ must have dimensions of mass (force/acceleration), and $q^2/(\epsilon_0 c^2 a_0)$ indeed has units of mass.

### A.6 Energy-Based Estimate of $C$

The scaling of the coefficient $C$ can be obtained more physically from the dependence of the electromagnetic self-energy of a charged sphere on its radius. For a spherical shell of radius $a$,

$$U_{em} = \frac{q^2}{8\pi\epsilon_0 a}.$$

Let the radius undergo a small deformation $a(t) = a_0 + \xi(t)$. Expanding to first order gives

$$U_{em}(a) \approx \frac{q^2}{8\pi\epsilon_0 a_0} - \frac{q^2}{8\pi\epsilon_0 a_0^2}\xi.$$

The electromagnetic force acting on the internal coordinate follows from $f_{em} = -\partial U_{em}/\partial \xi$. When the center of mass accelerates, retardation across the sphere introduces the time scale $a_0/c$. Expanding the retarded self-interaction to leading order in $a_0/c$ produces a term proportional to $\ddot{\mathbf{v}}$, yielding

$$F_{em}(t) = C\ddot{\mathbf{v}}(t), \quad C = \frac{q^2}{6\pi\epsilon_0 c^2 a_0}.$$

The numerical factor $1/(6\pi)$ arises from the angular integration of the retarded self-field over the spherical surface, as in the standard rigid-sphere self-force calculation.

### A.7 Summary

The linearized force on the breathing mode is therefore

$$f_{em}^{(linear)}(t) = C\ddot{\mathbf{v}}(t), \quad C = \frac{q^2}{6\pi\epsilon_0 c^2 a_0}.$$

This expression is used in Section 2.3 to derive the adiabatic response of the internal mode and the effective kernel. The derivation shows that $C$ is not a free parameter but is fixed by the same physical scales that govern the radiation reaction force on the center of mass. Its value follows from the geometry of the charge distribution and the finite speed of light, reinforcing the ESD philosophy that internal structure and causal propagation are inseparably linked.